\documentclass[conference]{IEEEtran}

\IEEEoverridecommandlockouts
\usepackage{amsmath}
\usepackage{graphicx}
\usepackage{tabularx}
\usepackage{amssymb} 
\usepackage{gensymb}
\usepackage{multirow}
\usepackage{subfig}
\usepackage{booktabs}
\usepackage[table]{xcolor}
\usepackage{amsmath}
\usepackage{array}
\usepackage{bm}
\usepackage{nicefrac}
\newcolumntype{C}[1]{>{\centering\arraybackslash}m{#1}}

\usepackage{tikz}
\usetikzlibrary{shapes}

\usetikzlibrary{plotmarks}
\newcommand\marksymbol[2]{\tikz[#2,scale=1.2]\pgfuseplotmark{#1};}

\usepackage[bookmarks=false,colorlinks=true,citecolor=blue]{hyperref}

\begin{document}

\title{Detection of Gait Asymmetry \\ Using Indoor Doppler Radar}

\author{\IEEEauthorblockN{Ann-Kathrin Seifert, Abdelhak M. Zoubir}
	\IEEEauthorblockA{Signal Processing Group\\
		Technische Universit\"{a}t Darmstadt, 64283 Darmstadt, Germany\\
		Email: \{seifert, zoubir\}@spg.tu-darmstadt.de}
	\and
	\IEEEauthorblockN{Moeness G. Amin}
	\IEEEauthorblockA{Center for Advanced Communications\\
		Villanova University, Villanova, PA 19085 USA\\
		Email: moeness.amin@villanova.edu}
\thanks{\textcircled{c} 2019 IEEE. Personal use of this material is permitted. Permission from IEEE must be obtained for all other uses, in any current or future media, including reprinting/republishing this material for advertising or promotional purposes, creating new collective works, for resale or redistribution to servers or lists, or reuse of any	copyrighted component of this work in other works.	
This work is supported by the Alexander von Humboldt Foundation, Bonn, Germany.}	
}

\maketitle
 
\begin{abstract}
Doppler radar systems enable unobtrusive and privacy-preserving	long-term monitoring of human motions indoors. In particular, a person's gait can provide important information about their state of health. Utilizing micro-Doppler signatures, we show that radar is capable of detecting small differences between the step motions of the two legs, which results in asymmetric gait. Image-based and physical features are extracted from the radar return signals of several individuals, including four persons with different diagnosed gait disorders. It is shown that gait asymmetry is correctly detected with high probability, irrespective of the underlying pathology, for at least one motion direction.
\end{abstract}

\IEEEpeerreviewmaketitle

\section{Introduction}

Gait analysis plays a key role in many areas such as medical diagnosis, rehabilitation and fall risk prediction \cite{Mur14,Pir17}. Early detection of balance or gait problems enables timely diagnosis of many neurological, orthopedic and medical conditions \cite{Pir17}. Further, changes in gait patterns can be precursors of falls \cite{Pir17}, which are the most frequent cause of severe injuries in the elderly aged over 65 years \cite{Pir17} and the second leading cause of death for that age group \cite{WHO07}.

For medical gait analysis, an important domain to be analyzed is the asymmetry of the gait, which refers to the differences between the left and right leg's motions \cite{Gou17,Sel14}. Many pathologies lead to gait asymmetry with various degrees. On the other hand, in rehabilitation, professionals work toward re-establishing a symmetric gait of the patient. Thus, detecting gait asymmetry provides useful information to clinicians and can help in identifying the onsets of many pathological disorders or assess the state of rehabilitation.

Most of recent mobile gait analysis systems are based on wearable sensors (for a review see e.g.~\cite{Mur14,Che16}). However, they can be intrusive and uncomfortable since they need to be worn on the body. The capabilities of non-wearable sensors, such as video cameras, are often restrained due to privacy concerns. For these reasons, we propose using Doppler radar systems for long-term in-home gait analysis. Radar provides an unobtrusive and privacy-preserving sensing of human motions. It can monitor the gait remotely from the distance, and thus does not hinder normal movement, nor alter the person's daily routine. Further, radar can operate in all lighting conditions and is insensitive to clothing. Thus, it represents a fast, cost-efficient and safe supplement to clinical gait analysis, which is often time-consuming, costly and lacks reproducibility \cite{Sim04}.   

Electromagnetic sensing of human activities has become of increased interest over the past years \cite{Ami17,Che14}. More recently, the so-called radar micro-Doppler signatures have been used to study and assess gait motions in more detail, e.g.,~for identifying persons \cite{Van18,Tek18} or to perform gait recognition \cite{Sey18,Gur17,Wan14,Sei17a,Sei18,Sei18a,Sei18b}. Detection of abnormal micro-Doppler step signatures has previously been investigated in \cite{Sei17}. While most research work is based on predefined classes (e.g. normal vs.~assisted gait or normal vs.~abnormal steps), this assumption is very restrictive in practice.

In this work, we aim to further the understanding of micro-Doppler signatures of human gait and validate the idea of using radar for medical applications. We model the probability of observing an asymmetric gait using a combination of image-based and kinematically related features of the radar returns. A subsequent detector is designed to have a false alarm rate of maximal 5\,\%. Based on real radar data, including data of four test subjects with different diagnosed gait disorders, asymmetric gait is correctly identified with a probability of up to 100\,\% for at least one of the considered motion directions. 

The remainder of the paper is structured as follows. Sec.~\ref{sec:mDsignatures} presents the micro-Doppler gait signatures of different individuals with and without gait disorders. Next, Sec.~\ref{sec:features} outlines the processing steps to obtain descriptive features from the radar return signals, and introduces the model for gait asymmetry detection. Sec.~\ref{sec:results} gives the experimental results based on real radar data and Sec.~\ref{sec:conclusion} concludes the paper. 

\section{Micro-Doppler Signatures of Human Gait}
\label{sec:mDsignatures}
Since the back-scattered radar returns from human motions are highly non-stationary and contain multiple components of different body parts, they are most often analyzed using time-frequency representations. In the Doppler frequency vs.~time domain, the time-varying Doppler shifts corresponding to each signal component can be revealed. Micro-motions, such as swinging arms or legs, lead to so-called micro-Doppler shifts \cite{Che11}. In this way, different motions result in different micro-Doppler signatures. Note that, at a given time instant, the observed Doppler frequencies describe the velocities, accelerations, and higher order motion moments of the corresponding body parts.

Typically, the spectrogram is used to represent human micro-Doppler signatures in the joint-variable domain \cite{Che14,Ami17}. For a discrete-time signal ${s}(n)$ of length $N$, the spectrogram is defined as the squared magnitude of the short-time Fourier transform (STFT) \cite{Opp99}
\vspace{-0.3em}
\begin{equation}
\mathrm{S}(n,k) = \left| \sum_{m=0}^{L-1} w(m) s(n+m) \exp{\left(-j 2 \pi \frac{mk}{K} \right)} \right|^2, 
\label{eq:spectrogram}
\end{equation}
for $n = 0, \dots, N-1$, where $L$ is the length of the smoothing window $w(\cdot)$, $k$ is the discrete frequency index with $k = 0, \dots, K-1$, and $N, L, K \in \mathbb{N}$.

\begin{figure}[!t]
	\vspace{-0.8em}
	\subfloat[Person A]{\includegraphics[clip, trim= 0 0 25 18, width=0.5\columnwidth]{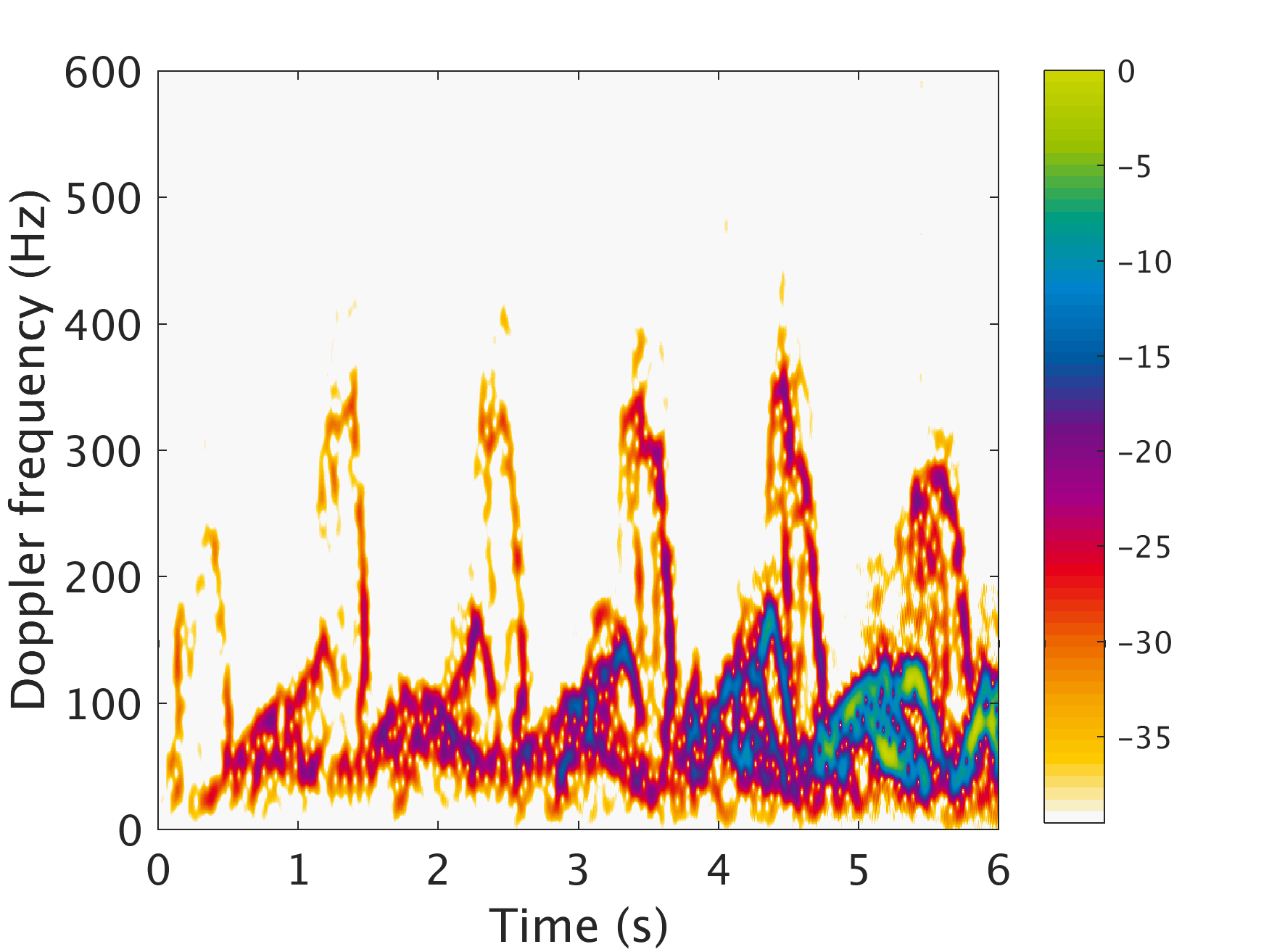}%
		\label{A}}
	\subfloat[Person A]{\includegraphics[clip, trim= 0 0 25 18, width=0.5\columnwidth]{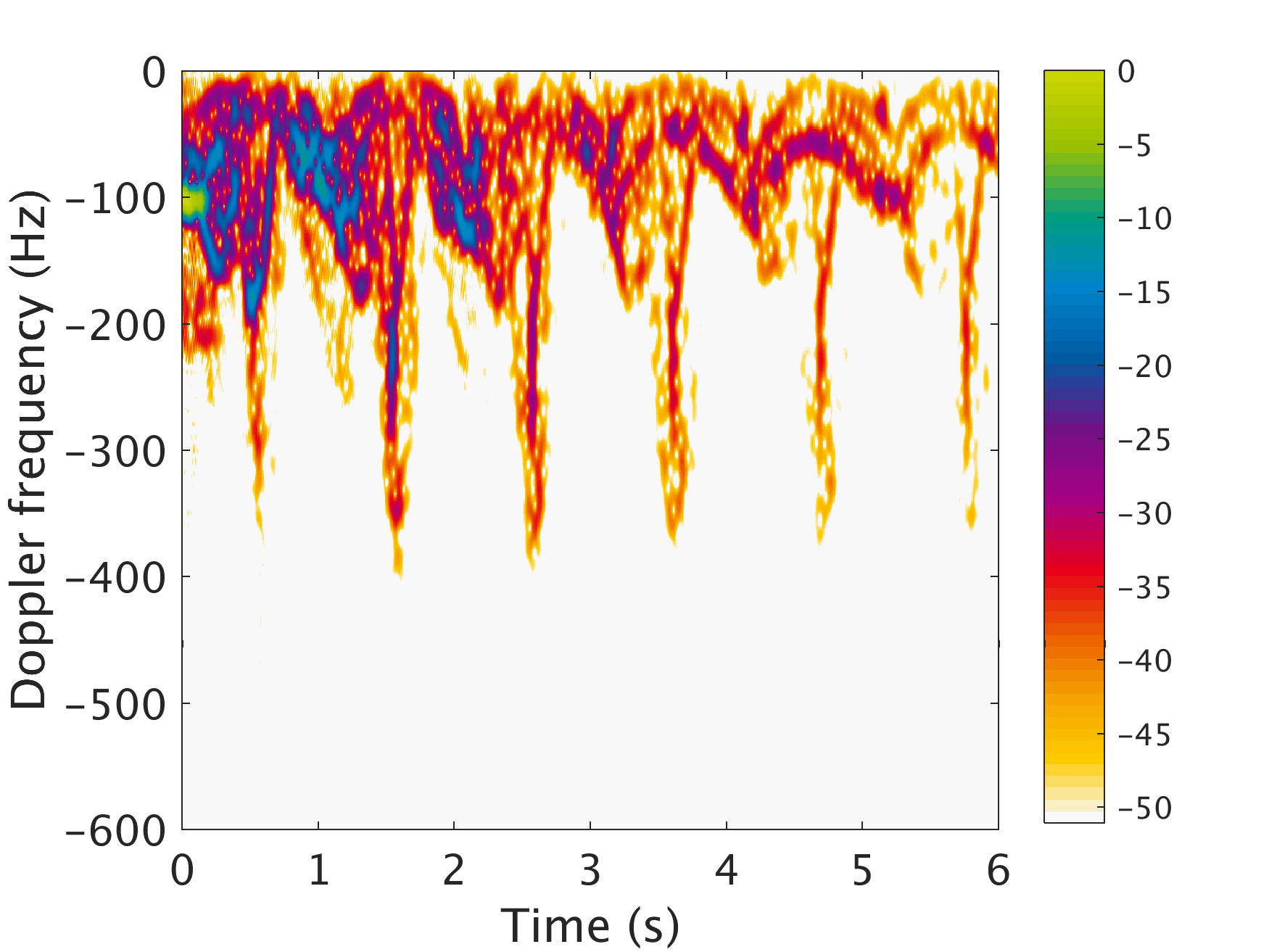}%
		\label{Aa}}
	\vspace{-0.8em}
	\subfloat[Person A - limping]{\includegraphics[clip, trim= 0 0 25 18, width=0.5\columnwidth]{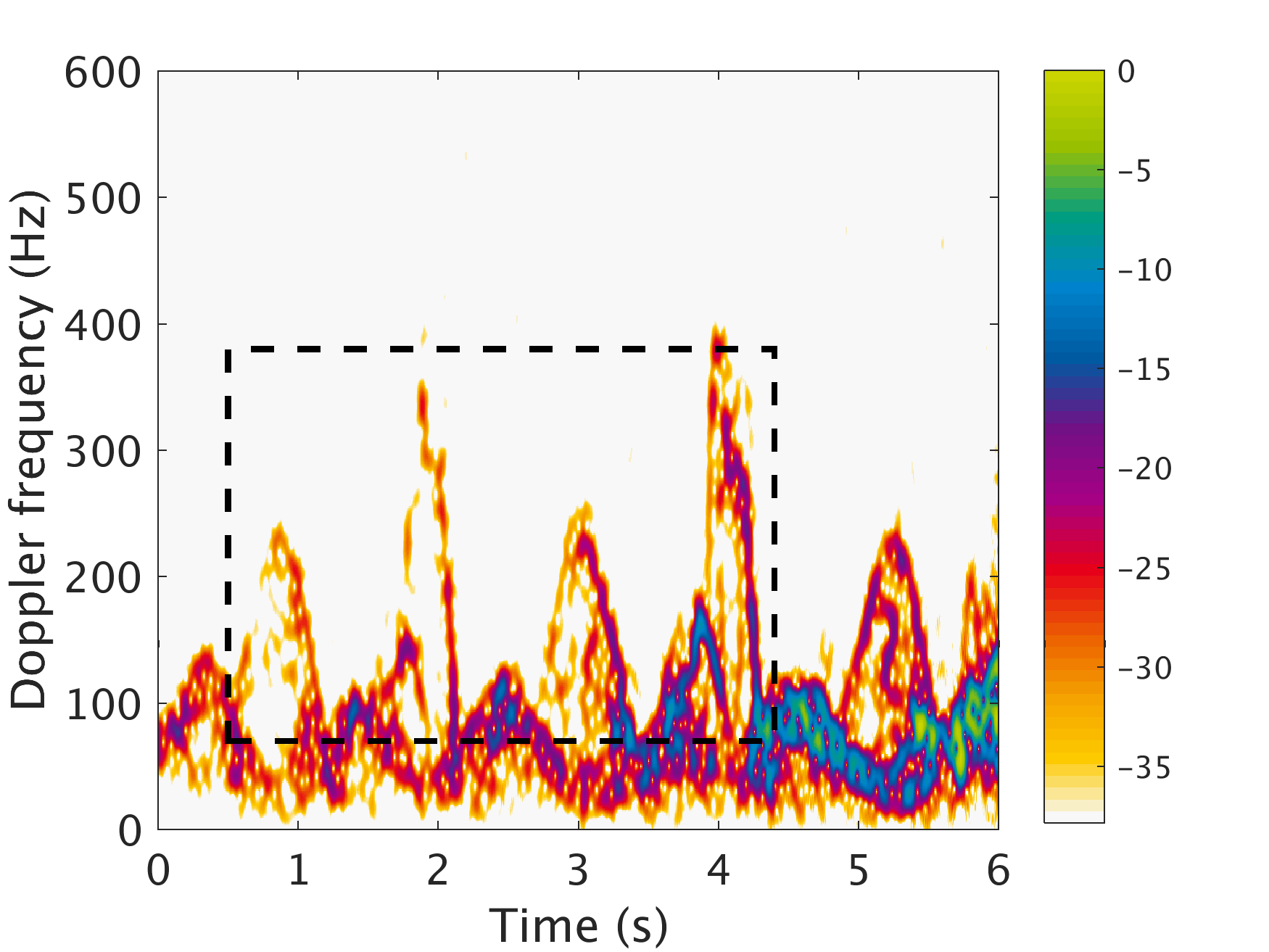}%
		\label{LA}}
	\subfloat[Person A - limping]{\includegraphics[clip, trim= 0 0 25 18, width=0.5\columnwidth]{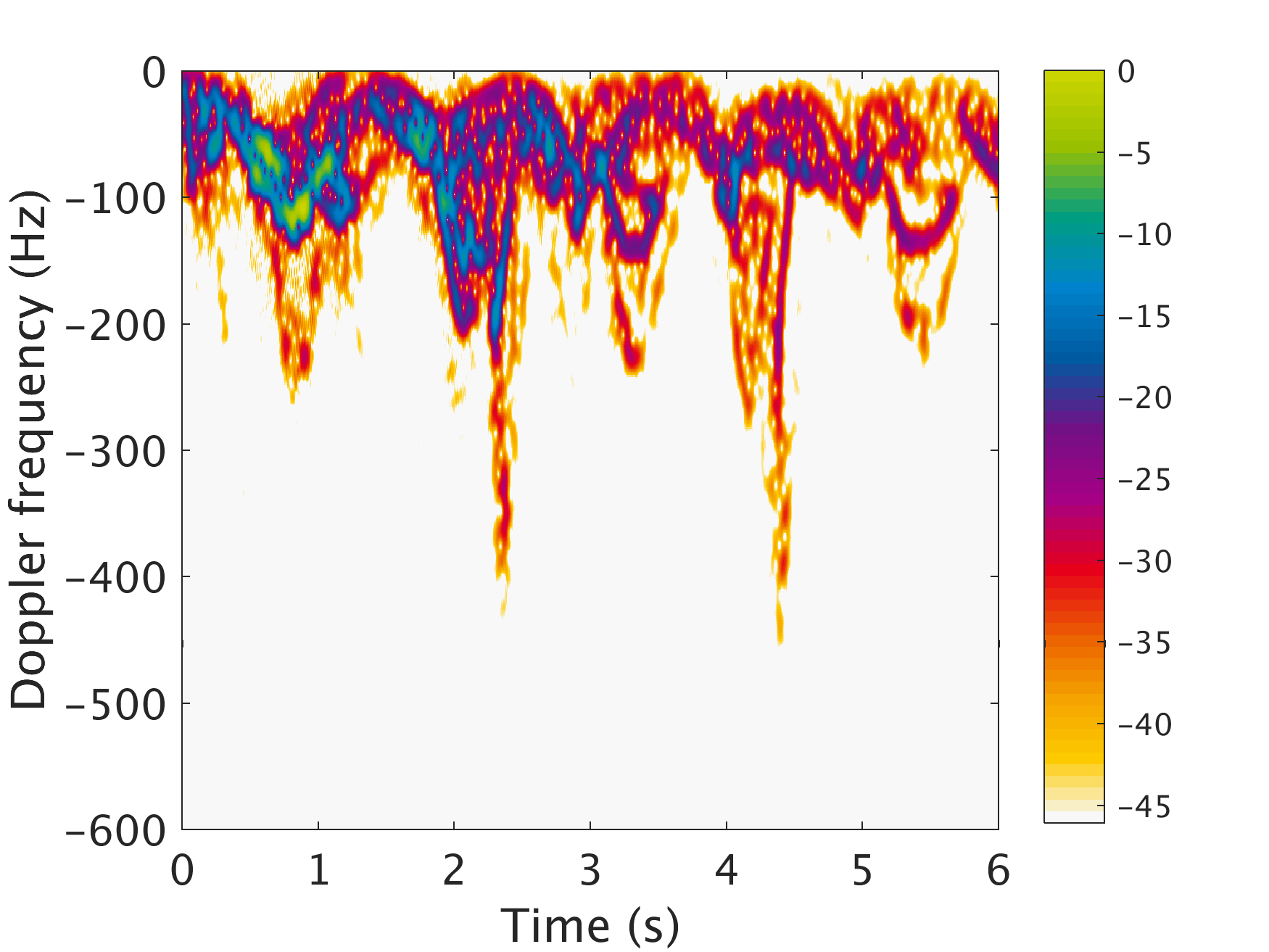}%
		\label{LAa}}
	\caption{Examples of spectrograms for a healthy individual moving toward (left) and away from (right) the radar system. In (c) and (d) the person simulated a limping gait. The color indicates the energy level in dB.}
	\label{fig:specs_A}
	\vspace{-0.8em}
\end{figure}

\begin{figure}[!t]
	\vspace{-0.8em}
	\subfloat[Person K]{\includegraphics[clip, trim= 0 0 25 18,width=0.5\columnwidth]{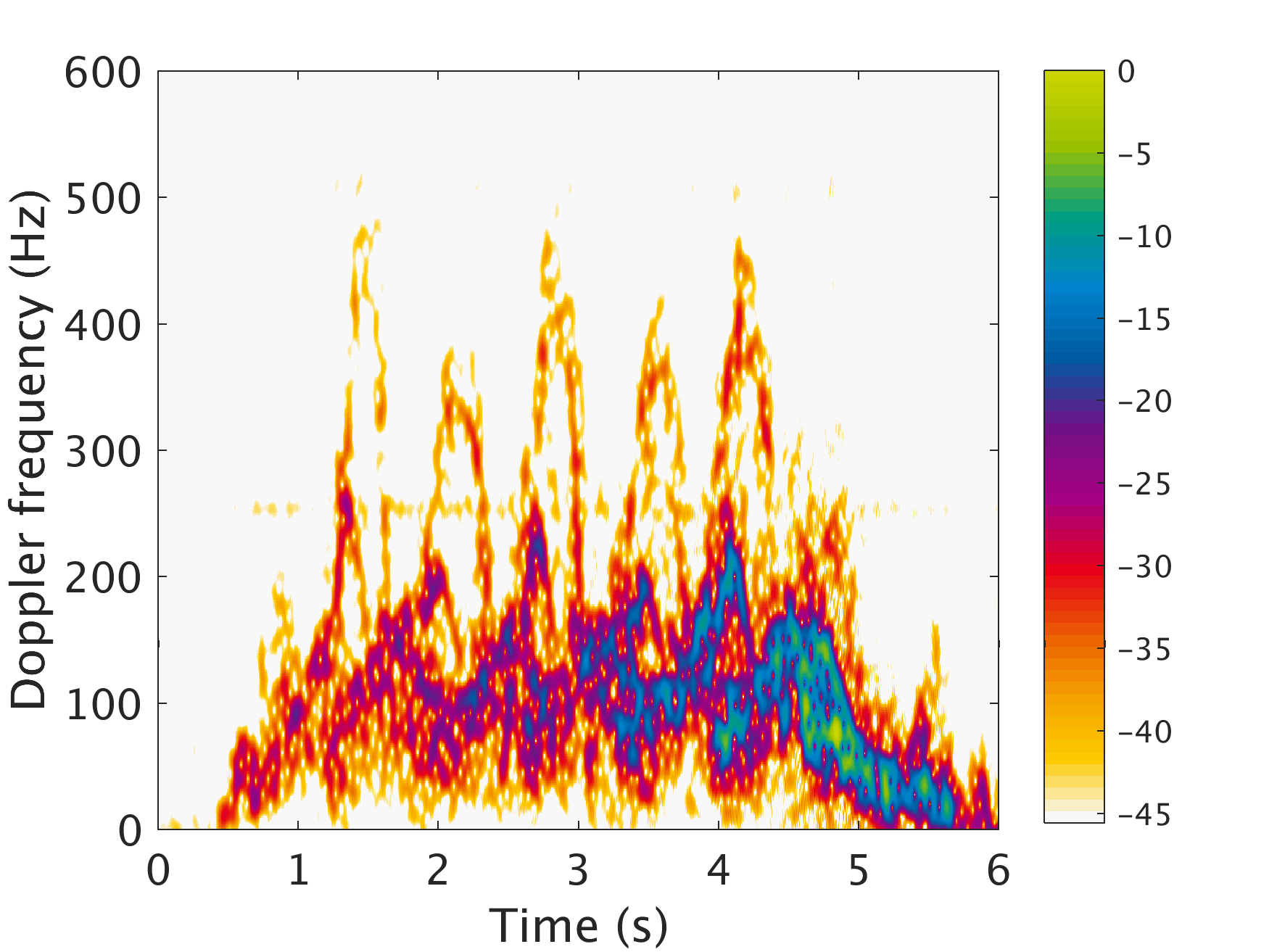}%
		\label{B}}
	\subfloat[Person K]{\includegraphics[clip, trim= 0 0 25 18,width=0.5\columnwidth]{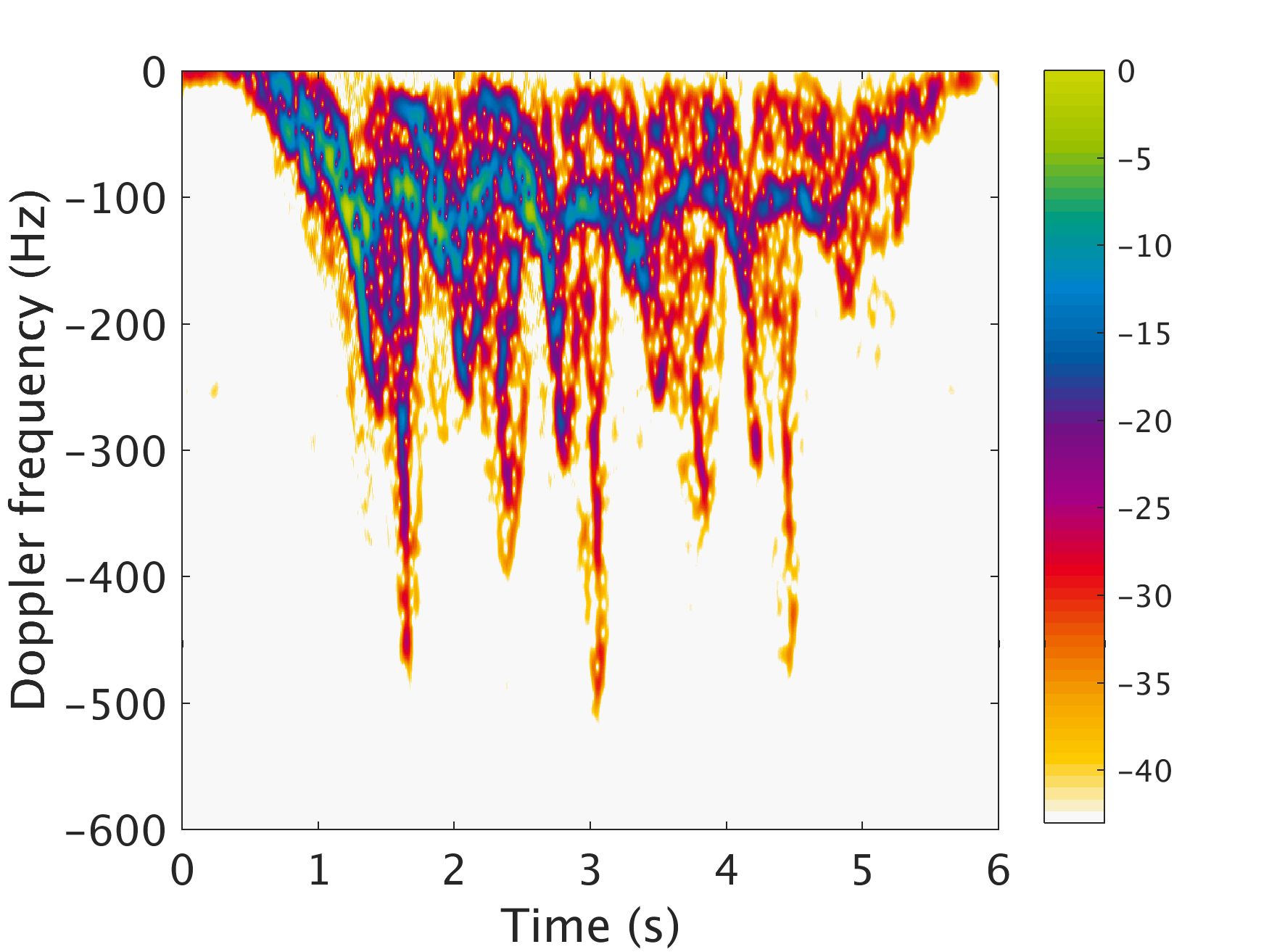}%
		\label{Ba}}
	\vspace{-0.8em}
	\subfloat[Person L]{\includegraphics[clip, trim= 0 0 25 18,width=0.5\columnwidth]{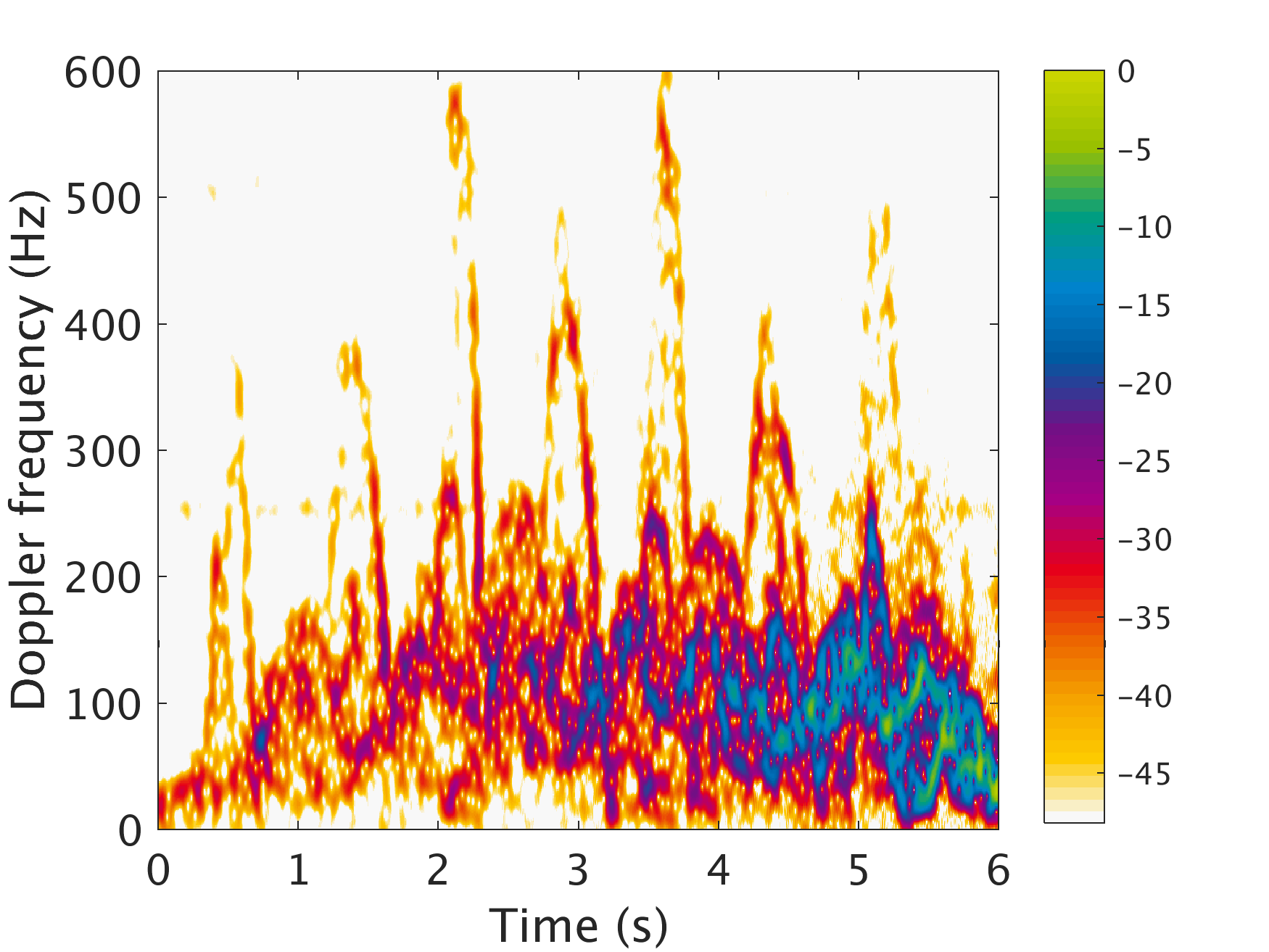}%
		\label{C}}
	\subfloat[Person L]{\includegraphics[clip, trim= 0 0 25 18,width=0.5\columnwidth]{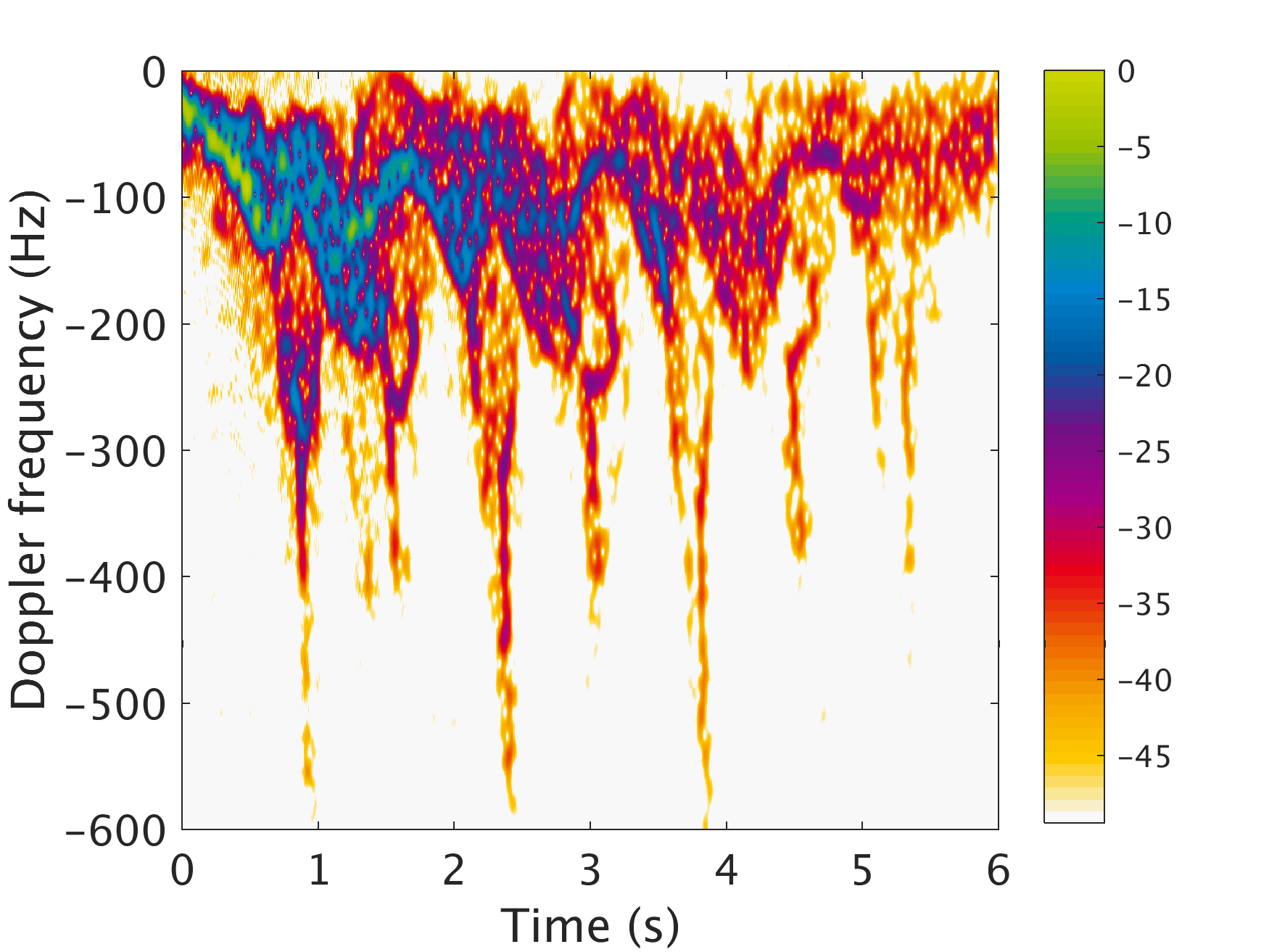}%
		\label{Ca}}
	\vspace{-0.8em}
	\subfloat[Person M]{\includegraphics[clip, trim= 0 0 25 18, width=0.5\columnwidth]{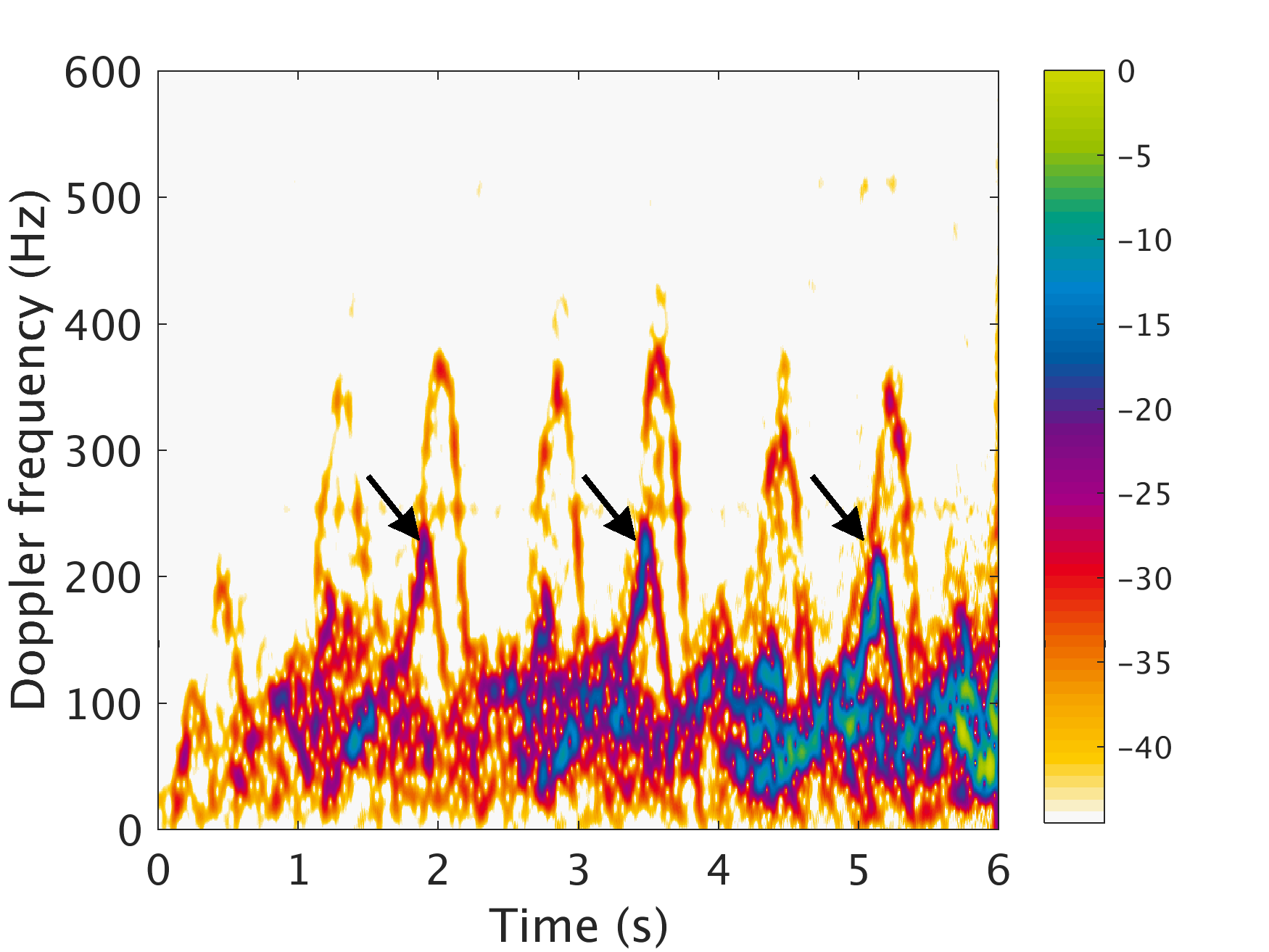}%
		\label{D}}
	\subfloat[Person M]{\includegraphics[clip, trim= 0 0 25 18, width=0.5\columnwidth]{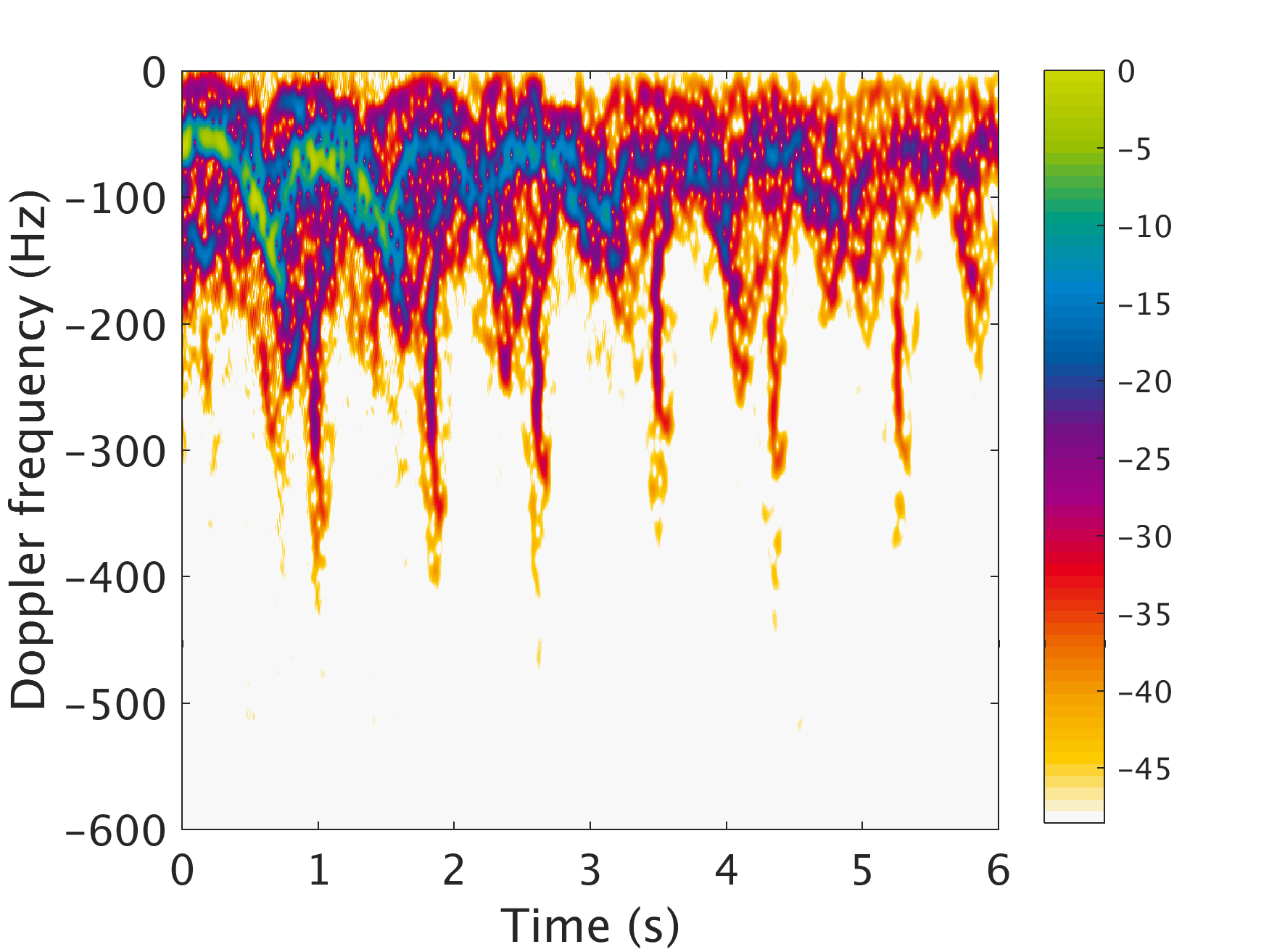}%
		\label{Da}}
	\vspace{-0.8em}
	\subfloat[Person N]{\includegraphics[clip, trim= 0 0 25 18, width=0.5\columnwidth]{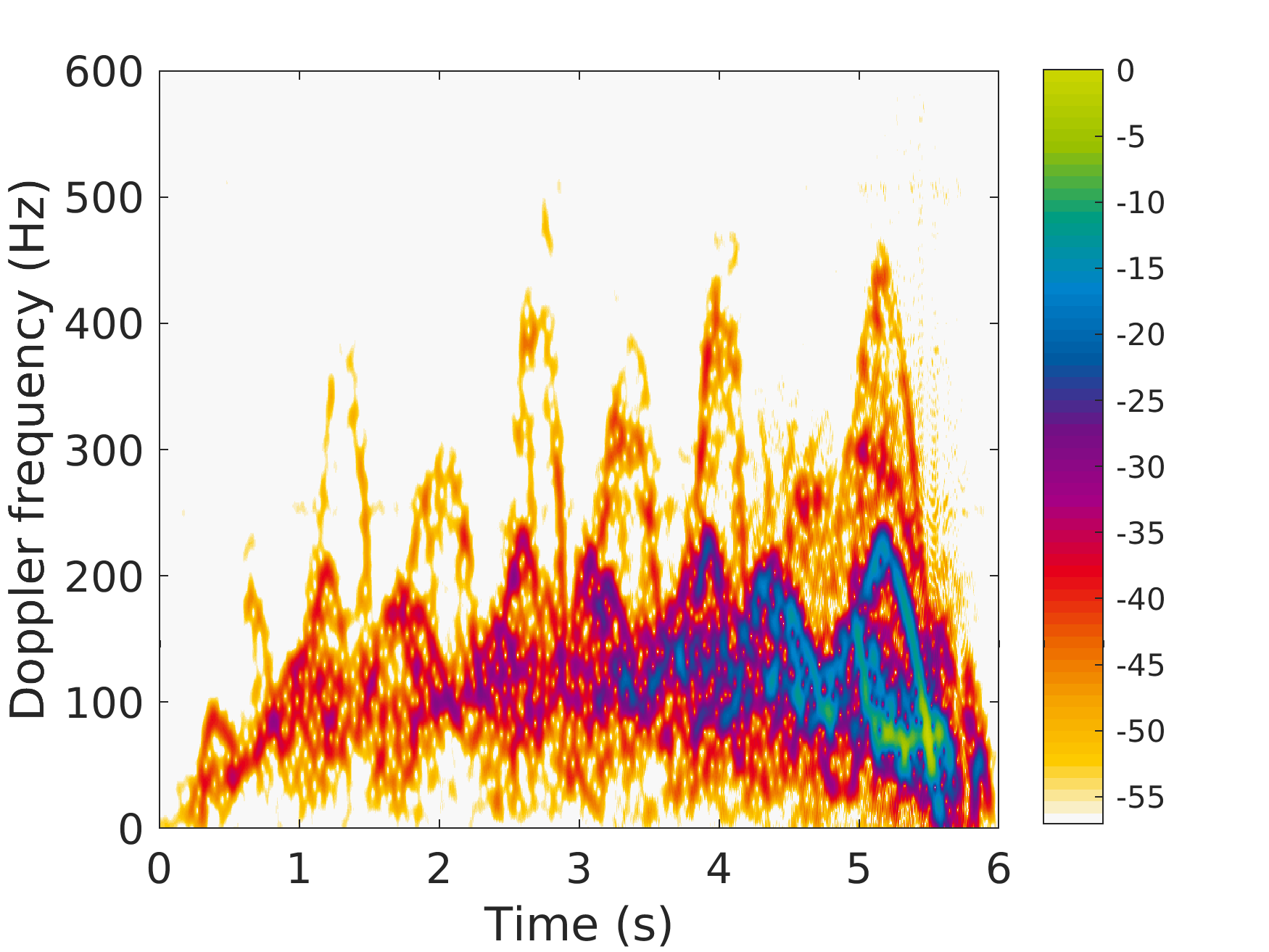}%
		\label{E}}
	\subfloat[Person N]{\includegraphics[clip, trim= 0 0 25 18, width=0.5\columnwidth]{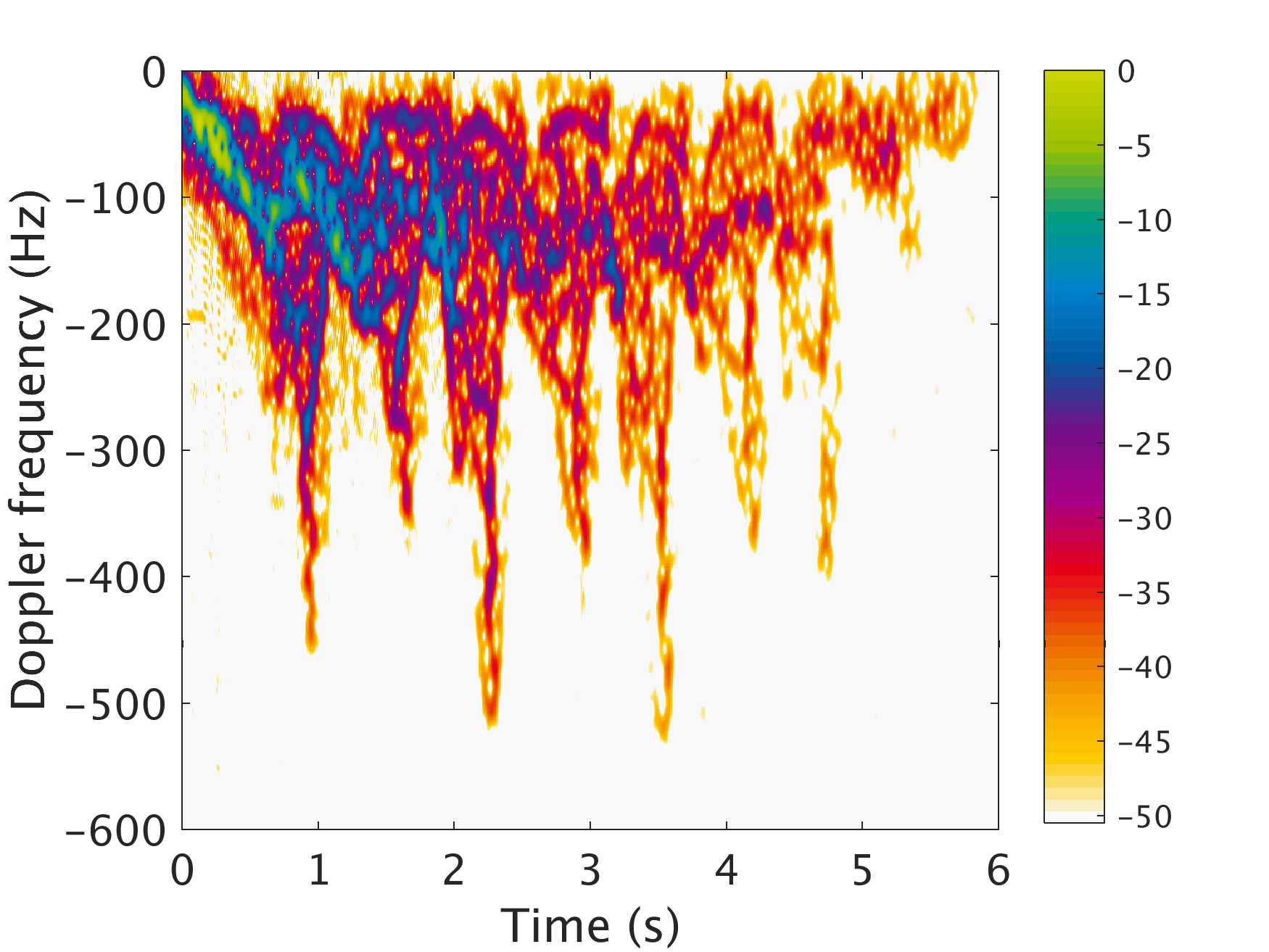}%
		\label{Ea}}
	\caption{Examples of spectrograms for four individuals with diagnosed gait disorders moving toward (left) and away from (right) the radar system. The color indicates the energy level in dB.}
	\label{fig:specs_rest}
	\vspace{-0.8em}
\end{figure}

Figs.~\ref{fig:specs_A} and \ref{fig:specs_rest} show examples of spectrograms of human gait performed toward (left column) and away from (right column) the radar system. Here, the micro-Doppler signatures of five different individuals are shown. Person A is a healthy subject, and Persons K--N have a pathological gait due to different reasons (for details see \cite{Sei18b}). Figs.~\ref{fig:specs_A}\subref{A}-\subref{Aa} depict the micro-Doppler signatures of Person A walking normally, whereas Figs.~\ref{fig:specs_A}\subref{LA}-\subref{LAa} show the same person performing a limping gait. In this case, limping was simulated by not bending one of the knees while walking. The radar micro-Doppler signatures in Figs.~\ref{fig:specs_rest}\subref{B}-\subref{Ea} clearly expose the asymmetry in the gait, since every other micro-Doppler step signature is different. Typically, one of the swinging legs reveals a lower Doppler shift, meaning it swings with a lower radial velocity. This behavior can result from a decreased stance phase for one leg, e.g., in an attempt to keep the load to a minimum, which is compensated by a longer swing phase. However, in the case of Person M the gait asymmetry reveals itself in the knee's motions (see arrows in Fig.~\ref{fig:specs_rest}\subref{D}), rather than in varying maximal Doppler shifts of the feet. This indicates that different gait disorders lead to distinct radar micro-Doppler signatures, which can be key to future in-home gait monitoring systems.

\begin{figure*}[!t]
	\centering
	\includegraphics[clip, trim= 15 0 0 0, width=0.95\linewidth]{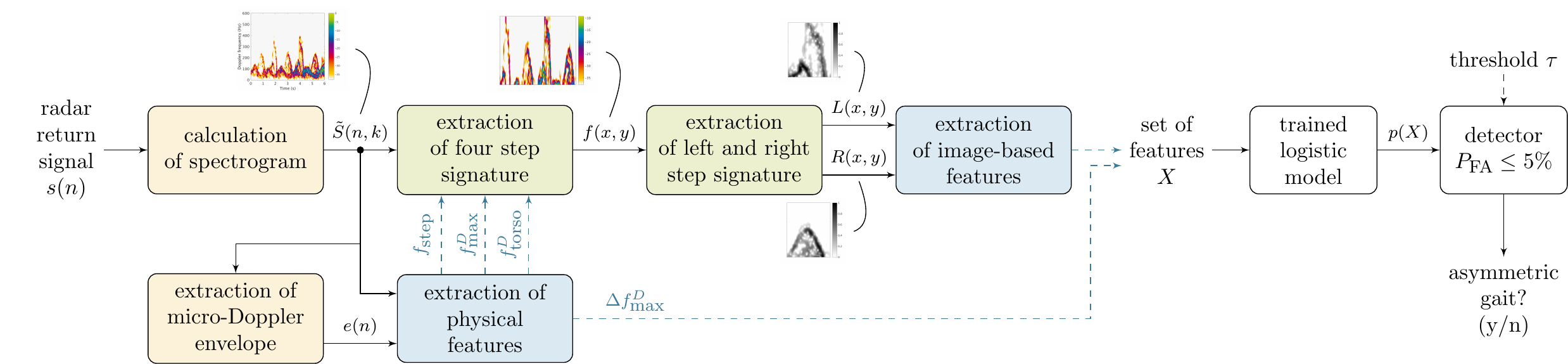}
	\caption{Overview of processing steps to detect asymmetric gait from a radar return signal: preprocessing (yellow), micro-Doppler step signature extraction (green), feature extraction (blue), and detection (white).}
	\label{fig:flow_diag}
	\vspace{-0.5em}
\end{figure*}

\section{Feature Extraction and Model for Gait Asymmetry Detection}
\label{sec:features}
Fig.~\ref{fig:flow_diag} provides an overview of the processing steps to obtain salient features for gait asymmetry detection from a radar return signal. The obtained features are then used to model the probability of observing an asymmetric gait. The details are described in the following sections.

\subsection{Preprocessing}
The complex zero-mean radar return signal ${s}(n)$ is processed to obtain the spectrogram according to (\ref{eq:spectrogram}), where a Hamming window of length $L = 255$ and $K = 2048$ discrete frequency bins are used. The signal length assumes $N=f_s \cdot T = 2560\,\text{Hz} \cdot 6\,\text{s} = 15360$ samples, where $f_s$ is the sampling frequency and $T$ is the measurement duration. After generating the spectrogram, an adaptive thresholding method is applied to suppress the background noise in the time-frequency domain \cite{Kim09}. From the noise-reduced spectrogram $\tilde{S}(n,k)$, we calculate the envelope of the micro-Doppler signatures, which is used to estimate the step rate $f_\text{step}$ and the maximal Doppler shift $f^D_\text{max}$ (for details see \cite{Sei18}). Note that, the latter corresponds to the maximal swing velocity of the feet during walking. Further, we determine the average Doppler shift of the torso $f^D_\text{torso}$ \cite{Sei18}, which describes the average walking speed of the person. Utilizing the above information, we automatically extract a representative portion of the spectrogram of four micro-Doppler step signatures $f(x,y)$ of size $M_x \times M_y$, as indicated by the dashed box in Fig.~\ref{fig:specs_A}\subref{LA}.

\begin{figure}[!t]
	\centering
	\vspace{-1em}
	\subfloat[$L(x,y)$]{\includegraphics[clip, trim= 15 15 15 18, width=0.4\columnwidth]{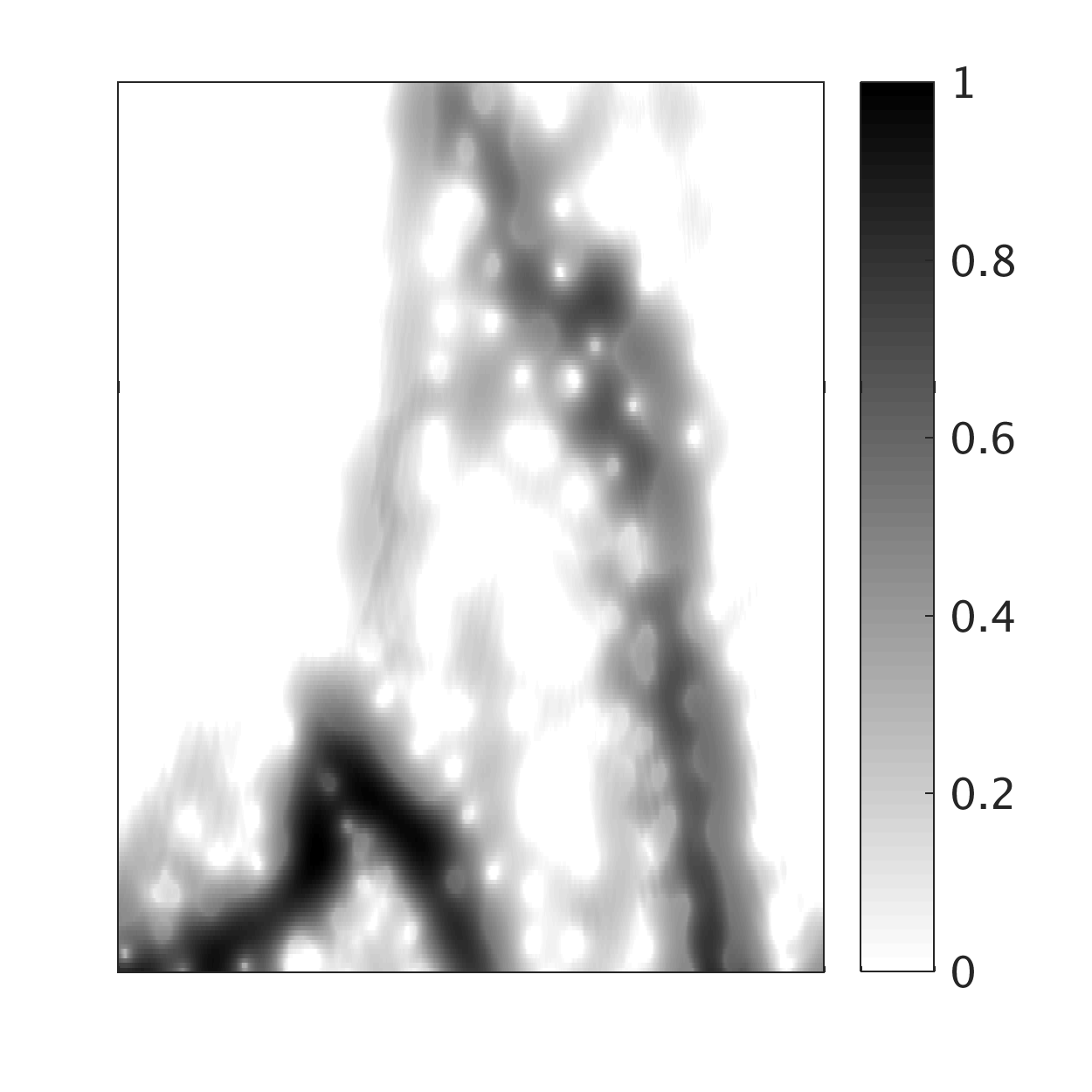}%
		\label{L_step}}
	\hfil
	\subfloat[$R(x,y)$]{\includegraphics[clip, trim= 15 15 15 18, width=0.4\columnwidth]{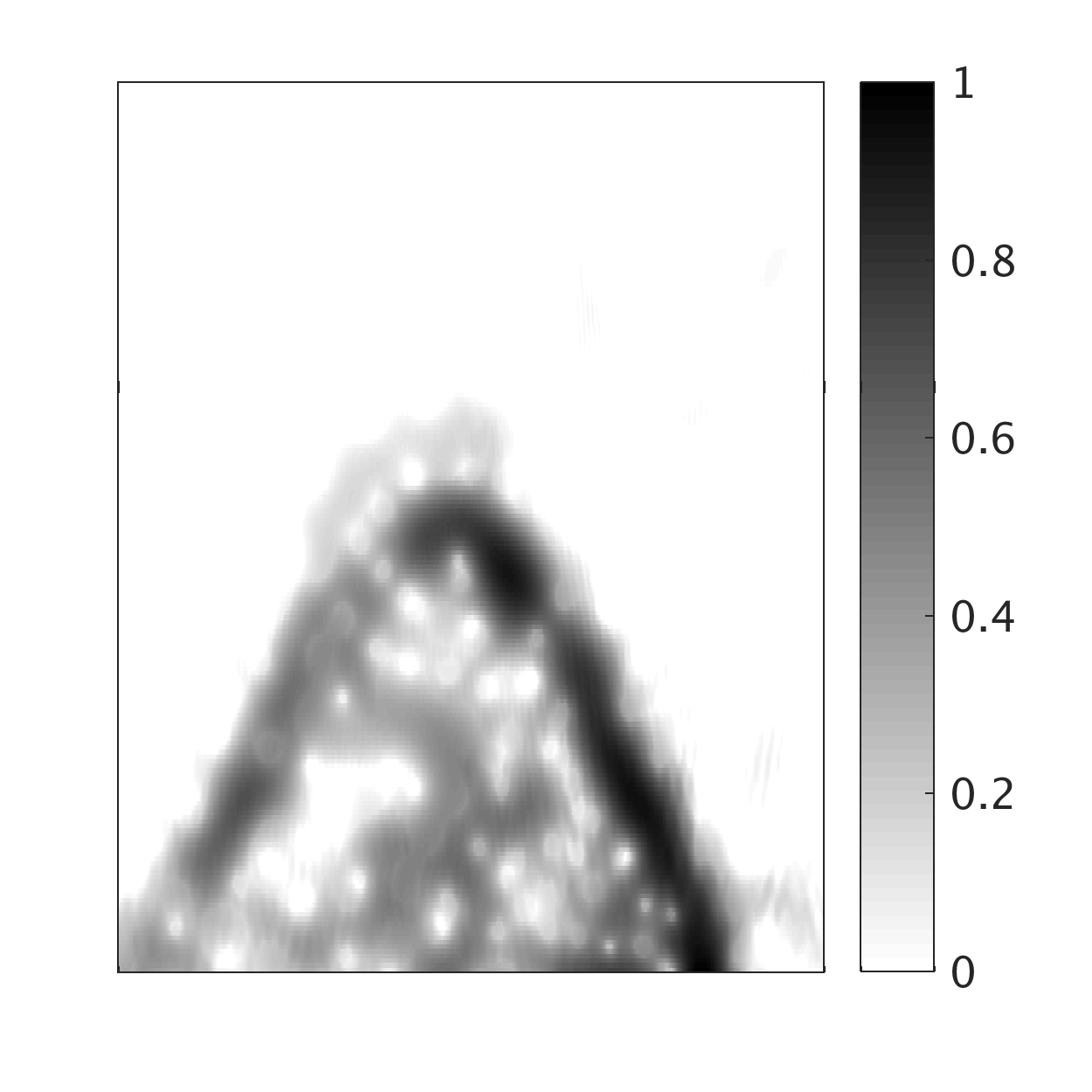}%
		\label{R_step}}
	\caption{Average (a) left and (b) right micro-Doppler step signatures extracted from the spectrogram in Fig.~\ref{fig:specs_A}\,(c).}
	\label{fig:avg_steps}
	\vspace{-1em}
\end{figure}

\subsection{Extraction of Micro-Doppler Step Signatures}
In order to calculate an average micro-Doppler step signature of the left and right leg separately, we first estimate the time locations of the steps in $f(x,y)$ by finding the maxima in the envelope signal. Then, four individual step signatures of size $N_x \times N_y$ are extracted, where $N_x = \nicefrac{2}{3} \cdot f_s / f_\text{step} < M_x$ and $N_y = M_y$. The latter relates to the number of frequency bins that correspond to the range between $1.5 \cdot f^D_\text{torso}$ and $f^D_\text{max}$. Next, every second step signature is averaged to yield a pair of average micro-Doppler step signatures. Since the envelope's peaks are susceptible to background noise, we refine the initial step time estimates by using an image registration technique. For this, the normalized 2D cross-correlation is calculated as 
\vspace{-0.5em}
\begin{multline}
\gamma (u,v) = \\
\frac{\sum_{x,y} \left[ f(x,y) - \bar{f}_{u,v} \right] \left[ t(x-u,y-v)- \bar{t}~\right]} { \sqrt{ \sum_{x,y} \left[ f(x,y) - \bar{f}_{u,v} \right]^2 \sum_{x,y} \left[ t(x-u,y-v)- \bar{t}~\right]^2} },
\end{multline}
where $f(x,y)$ is the spectrogram of four strides, $t(x,y)$ is an average micro-Doppler step signature  positioned at $(u,v)$, $\bar{f}_{u,v}$ is the mean of $f(x,y)$ in the region under $t$, and $\bar{t}$ is the mean of $t$ \cite{Lew95}. Since we are determined to find the maximal correlation in $x$ direction, ${u=0,\dots,M_x-N_x}$ and $v=0$, i.e., $t$ is not shifted in $y$-direction. 
The locations in $x$-direction that maximize $\gamma$ determine the new time instants of the steps. We perform this procedure for the left and right steps individually, and average the newly obtained signatures again. For easier notation, we hereafter use $R(x,y)$ and $L(x,y)$ to refer to the right and left leg's micro-Doppler signature, respectively. Note, however, that without any prior knowledge, we cannot infer which signature belongs to which leg. For further processing, the step signatures are converted to gray-scale images. Examples of average micro-Doppler step signatures are shown in Fig.~\ref{fig:avg_steps}. 

\subsection{Feature extraction}
Given the micro-Doppler step signatures $R(x,y)$ and $L(x,y)$ of size $N_x \times N_y$, we aim to quantify the (dis)similarity between them. The intuition being that the more similar $R(x,y)$ and $L(x,y)$ are, the more symmetric the gait. To this end the salient features related to motion kinematics listed in Table~\ref{tab:features} are extracted.

\begin{table}[!t]
	\caption{Image-based and physical features of micro-Doppler step signatures used for gait asymmetry detection.}
	\centering
	\begin{tabular}{l l}
		\toprule
		Feature & Symbol \\ 
		\cmidrule{1-2}
		correlation coefficient & $r$ \\ 
		\medspace -- at high Doppler frequencies & $r_H$ \\ 
		\medspace -- at medium Doppler frequencies & $r_M$ \\ 
		\medspace -- at low Doppler frequencies & $r_L$ \\ 
		mean squared error & \text{MSE} \\ 
		mean absolute error & \text{MAE} \\ 
		mean structural similarity index  & \text{MSSIM} \\ 
		difference of maximal Doppler shifts & $\Delta f^D_\text{max}$ \\
		\bottomrule
	\end{tabular}
	\label{tab:features}
	\vspace{-0.8em}
\end{table}

\subsubsection{Correlation coefficient}
The correlation coefficient is calculated as \cite{Mit10}
\begin{equation}
r =  \frac{\sum_{x,y} \left[ L(x,y) - \bar{L}~\right] \left[ R(x,y) - \bar{R}~\right]} { \sqrt{ \sum_{x,y} \left[ L(x,y) - \bar{L}~\right]^2 \sum_{x,y} \left[ R(x,y) - \bar{R}~\right]^2} },
\end{equation}
where $\bar{L}$ and $\bar{R}$ are the mean of $L(x,y)$ and $R(x,y)$, respectively.

\subsubsection{Mean Squared Error (MSE)}
The mean square error (MSE) is given by \cite{Mit10}
\begin{equation}
\text{MSE} = \frac{1}{N_x N_y} \sum_{x,y} \left[ L(x,y) - R(x,y) \right]^2.
\end{equation}

\subsubsection{Mean Absolute Error (MAE)}
The mean absolute error (MAE)  is calculated as \cite{Mit10}
\begin{equation}
\text{MAE} = \frac{1}{N_x N_y} \sum_{x,y} \left| L(x,y) - R(x,y) \right| 
\end{equation}

\subsubsection{Mean Structural Similarity Index (MSSIM)}
The structural similarity (SSIM) index is defined as \cite{Wan04}
\begin{equation}
\text{SSIM}(R,L) =  \left[ l(R,L) \right]^\alpha \cdot \left[ c(R,L) \right]^\beta \cdot \left[ s(R,L) \right]^\gamma, 
\end{equation}
where $l(R,L)$, $c(R,L)$, and $s(R,L)$ refer to luminance, contrast and structural measures, and $\alpha = \beta = \gamma = 1$.
To obtain a single overall quality score of the images, we calculate the mean SSIM (MSSIM) given by
\begin{equation}
\text{MSSIM} = \frac{1}{N_x N_y} \sum_{x,y} \text{SSIM}(R,L).
\end{equation}

\subsubsection{Offset in maximal Doppler shifts}
Finally, we also consider a physical feature, i.e., a characteristic that can easily be interpreted. As mentioned in Sec.~\ref{sec:mDsignatures}, most asymmetric gaits can be identified by different maximal Doppler shifts of the step signatures. As such, we expect an asymmetric gait to have alternating high and low maximal Doppler shifts due to the steps. Thus, we calculate the average difference between the maximal Doppler shifts of the two legs, $\Delta f^D_\text{max}$. This is done by utilizing the detected peaks in the envelope signal, calculating the absolute differences in Doppler frequency of consecutive peaks, and averaging the result.

\subsection{Model for Asymmetry Detection in Gait}
\label{subsec:model}
In this work, we seek to model the probability of observing an asymmetric gait based on the extracted features from the previous section. That is, we attempt to answer the question: given a new measurement, how likely is it that we are observing an asymmetric gait? We model the probability of asymmetric gait $p(X)=\Pr(\textit{asymmetric gait}|X)$ using a logistic function given by \cite{Jam13}
\begin{equation}
p(X) = \frac{e^{b_0 + b_1 X_1 + \cdots + b_d X_d}}{1+e^{b_0 + b_1 X_1 + \cdots + b_d X_d}},
\label{eq:regmodel}
\end{equation}
where $X = (X_1,\dots,X_d)$ are $d$ features or predictors, and $b_0,\dots,b_d$ are the regression coefficients. The latter are estimated based on the training data using maximum likelihood estimation \cite{Jam13}. The final decision is based on comparing the probability of asymmetric gait against a threshold $\tau$, i.e., if $p(X) \ge \tau$ we decide for an asymmetric gait.

\section{Experimental results}
\label{sec:results}

\subsection{Experimental Setup}
\label{sec:expsetup}
The experimental radar data of ten healthy subjects (Persons A--J) were collected in an office environment at Technische Universit\"at Darmstadt, Germany. A 24\,GHz continuous-wave radar system \cite{Anc} was positioned at 1.15\,m above the floor, and the volunteers were asked to walk slowly back and forth between 1\,m and 4\,m in front of the radar. The gaits were performed in an 0$\degree$ angle relative to the radar's line-of-sight and, due to a slow walking speed, without major arm swinging. The same setup was used to record data of four additional subjects with diagnosed gait disorders (Persons K--N) at Villanova University, USA. The statistics of the participants and the number of measurements per person are given in Table~\ref{tab:exp_data}. In total, 471 samples are considered, out of which 271 measurements correspond to (simulated) asymmetric gait.

\begin{table}[!t]
	\caption{Statistics of participating test subjects.}
	\centering
	\begin{tabular}{l c c c c }
		\toprule
		\multirow{2}*{Person} & \multirow{2}*{Sex} & \multirow{2}*{Age} & Gait & \# of meas. \\ 
		 & & & disorder & toward / away\\
		\cmidrule{1-5}
		\multirow{2}*{A--J}  & \multirow{2}*{$8$ male, $2$ female} & \multirow{2}*{$23.8 \pm 2.6$} & no & 10 / 10 (each) \\
		& & & simulated & 10 / 10 (each) \\
		K  & female 			 & 	n.a.			& yes & 7 / 6\\	
		L  & female 			 &  n.a.		    & yes & 11 / 9\\		
		M  & female 			 & 	n.a.			& yes & 7 / 5 \\		
		N  & female 			 & 	n.a.			& yes & 13 / 13\\			
		\cmidrule{1-5}
		A--N & $8$ male, $6$ female & - & - & 238 / 233	\\	
		\bottomrule
	\end{tabular}
	\label{tab:exp_data}
	\vspace{-0.8em}
\end{table}

\subsection{Model Selection}
\begin{figure}
	\vspace{-0.5em}
	\centering
	\includegraphics[clip, trim= 0 0 25 18,width=0.85\columnwidth]{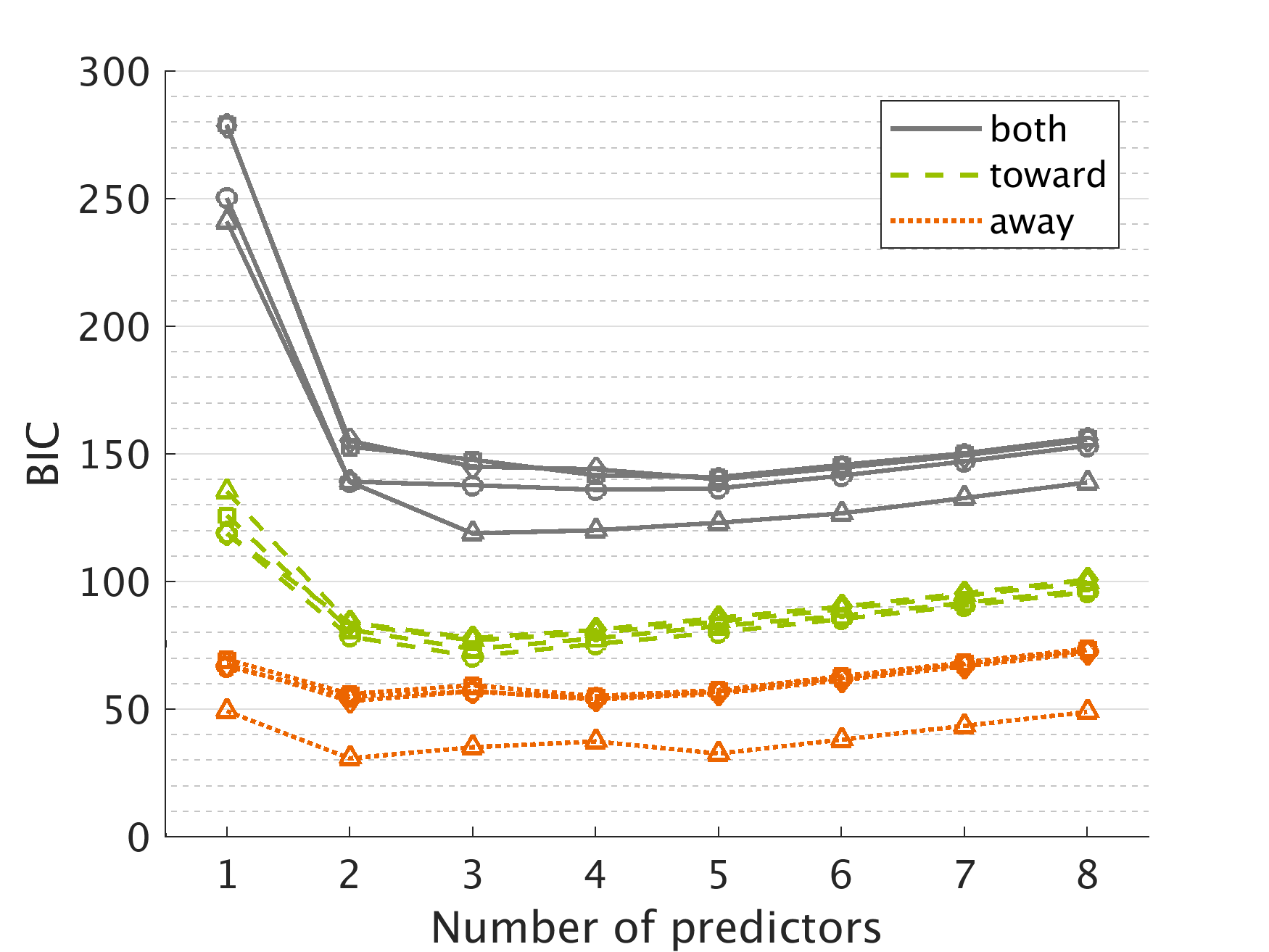}
	\caption{Bayesian Information Criterion (BIC) for model order selection based on available data excluding Person K (\protect\marksymbol{square}{black}), Person L (\protect\marksymbol{diamond}{black}), Person M (\protect\marksymbol{triangle}{black}), and Person N (\protect\marksymbol{o}{black}), respectively.}
	\label{fig:model_selection}
	\vspace{-0.8em}
\end{figure}

In order to compare models with different numbers of predictors, the Bayesian Information Criterion (BIC) \cite{Sch78} is utilized. Fig.~\ref{fig:model_selection} shows the lowest BIC values for each model order, i.e., models with $d$ predictors, $d = 1,\dots,8$, where all feature combinations were tested. For each scenario ('both', 'toward', and 'away'), the model which minimizes the BIC is chosen as the final model. The BIC is evaluated based on all available data excluding data of one of the four diagnosed test subjects at a time. As mentioned in Sec.~\ref{sec:mDsignatures}, Person M shows a slightly different gait asymmetry than the remaining subjects in the data set. Hence, excluding Person M from the model selection process, the BIC decreases for the scenarios 'both' and 'away', which indicates a better model fit. In particular, the BIC assumes significantly smaller values for all model orders compared to excluding the other three subjects when considering the 'away' scenario. This can be explained by the fact that identifying the gait asymmetry of Person M from behind is challenging (see Fig.~\ref{fig:specs_rest}\subref{Da}). 

In general, Fig.~\ref{fig:model_selection} suggests to model gait asymmetry separately for toward and away from radar motions, since the BIC curves for both, the separate 'toward' and 'away' scenario, lie below the curves of 'both'. Table~\ref{tab:best_models} lists the models which minimize the BIC for different training sets. Excluding data of one person at a time, the coefficient values along with the standard error and $p$-value for the corresponding model are given. The standard error describes how much the coefficient estimate on average deviates from the actual value. In this experiment, standard errors can be large due to the limited number of observations. A small $p$-value (typically $\le 0.05$) indicates that the predictor is associated with the response. We note that the included predictors are $r$, $r_L$, $r_H$, $\Delta f^D_\text{max}$ and MSSIM, while $r_M$, MSE and MAD are never selected. Remarkably, all models in the 'toward' scenario utilize the same predictors ($r_H$, $r_L$, $\Delta f^D_\text{max}$). For the 'away' scenario, the same models are obtained when excluding Person K or N ($r$, $r_L$, $\Delta f^D_\text{max}$, MSSIM), and Person L or M ($\Delta f^D_\text{max}$, MSSIM).

\begin{table}[!t]
	\caption{Selected predictors and estimated coefficients of the best logistic regression models for predicting the probability of asymmetric gait. Intercept refers to $b_0$ in (\ref{eq:regmodel}).}
	\centering
	\subfloat[Excluding Person K]{
		\begin{tabular}{l l r r r}
			\toprule
			 & Predictor & Coefficient & Std. error & $p$-value  \\ 
			\cmidrule{1-5}
			\multirow{4}{*}{\rotatebox[]{90}{toward}} & intercept  & $4.5905$ & $1.5588$ & $0.0032$ \\ 
			& $r_H$  					& $-7.9980$ 	& $2.4388$ 	& $0.0011$ \\ 
			& $r_L$  					& $-11.8010$ 	& $2.3603$ 	& $< 0.0001$ \\ 
			& $\Delta f^D_\text{max}$  	& $0.0588$ 		& $0.0236$ 	& $0.0127$ \\ 
			\cmidrule{1-5}
			\multirow{5}{*}{\rotatebox[]{90}{away}} & intercept & $3.2862 $ & $5.4349 $ & $0.5454$ \\ 
			& $r$  						& $29.6620$ 	& $11.3080$ 	& $0.0087$ \\ 
			& $r_L$ 		 			& $-20.8820$ 	& $7.6164$ 		& $0.0061$ \\ 
			& $\Delta f^D_\text{max}$  	& $0.2147$ 		& $0.0601$ 		& $0.0004$ \\ 
			& $\text{MSSIM}$ 		    & $-39.3310$ 	& $13.6330$ 	& $0.0039$ \\ 
			\bottomrule		
		\end{tabular}}
	
		\subfloat[Excluding Person L]{
		\begin{tabular}{l l r r r}
			\toprule
			& Predictor & Coefficient & Std. error & $p$-value  \\ 
			\cmidrule{1-5}
			\multirow{4}{*}{\rotatebox[]{90}{toward}} & intercept  & $3.7248$ & $1.3379$ & $0.0054$ \\ 
			& $r_H$  					& $-6.8083$ 	& $2.1841$ 	& $0.0018$ \\ 
			& $r_L$  					& $-10.9560$ 	& $2.0824$ 	& $< 0.0001$ \\ 
			& $\Delta f^D_\text{max}$  	& $0.0689$ 	    & $0.0235$ 	& $0.0034$ \\ 
			\cmidrule{1-5}
			\multirow{3}{*}{\rotatebox[]{90}{away}} & intercept & $10.5560$ & $5.0276$ & $0.0358$ \\ 
			& $\Delta f^D_\text{max}$  	& $0.1578$ 		& $0.0350$ 		& $< 0.0001$ \\ 
			& $\text{MSSIM}$ 		    & $-32.4720$ 	& $10.141$ 	& $0.0014$ \\ 
			\bottomrule				
		\end{tabular}}

		\subfloat[Excluding Person M]{
		\begin{tabular}{l l r r r}
		\toprule
		& Predictor & Coefficient & Std. error & $p$-value  \\ 
		\cmidrule{1-5}
		\multirow{4}{*}{\rotatebox[]{90}{toward}} & intercept  & $3.3993$ & $1.3488$ & $0.0117$ \\ 
		& $r_H$  					& $-6.9438$ 	& $2.1610$ 		& $0.0013$ \\ 
		& $r_L$  					& $-10.4210$ 	& $2.1297$ 	& $< 0.0001$ \\ 
		& $\Delta f^D_\text{max}$  	& $0.0729$ 		& $0.0231$ 		& $0.0016$ \\ 
		\cmidrule{1-5}
		\multirow{3}{*}{\rotatebox[]{90}{away}} & intercept & $24.1940$ & $13.3100$ & $0.0691$ \\ 
		& $\Delta f^D_\text{max}$  	& $0.32045$ 	& $0.1236$ 		& $0.0095$ \\ 
		& $\text{MSSIM}$ 		    & $-72.8930$ 	& $33.7520$ 	& $0.0308$ \\ 
		\bottomrule			
		\end{tabular}}

		\subfloat[Excluding Person N]{
		\begin{tabular}{l l r r r}
		\toprule
		& Predictor & Coefficient & Std. error & $p$-value  \\ 
		\cmidrule{1-5}
		\multirow{4}{*}{\rotatebox[]{90}{toward}} & intercept  & $3.6100$ & $1.3407$ & $0.0071$ \\ 
		& $r_H$  					& $-6.7969$ 	& $2.1803$ 	& $0.0018$ \\ 
		& $r_L$  					& $-10.8120$ 	& $2.0826$ 	& $< 0.0001$ \\ 
		& $\Delta f^D_\text{max}$  	& $0.0704$ 		& $0.0234$ 	& $0.0026$ \\ 
		\cmidrule{1-5}
		\multirow{5}{*}{\rotatebox[]{90}{away}} & intercept & $2.2173$ & $5.6481$ & $0.6946$ \\ 
		& $r$  						& $28.3000$ 	& $11.3790$ 	& $0.0129$ \\ 
		& $r_L$ 		 			& $-21.2280$ 	& $7.8985$ 		& $0.0072$ \\ 
		& $\Delta f^D_\text{max}$  	& $0.2161$ 		& $0.0601$ 		& $0.0003$ \\ 
		& $\text{MSSIM}$ 		    & $-35.5050$ 	& $13.6080$ 	& $0.0090$ \\ 
		\bottomrule		
	\end{tabular}}
	\label{tab:best_models}
	\vspace{-1.8em}
\end{table}

\subsection{Gait Asymmetry Prediction}
\begin{figure}
	\centering
	\includegraphics[clip, trim= 0 0 25 18,width=0.85\columnwidth]{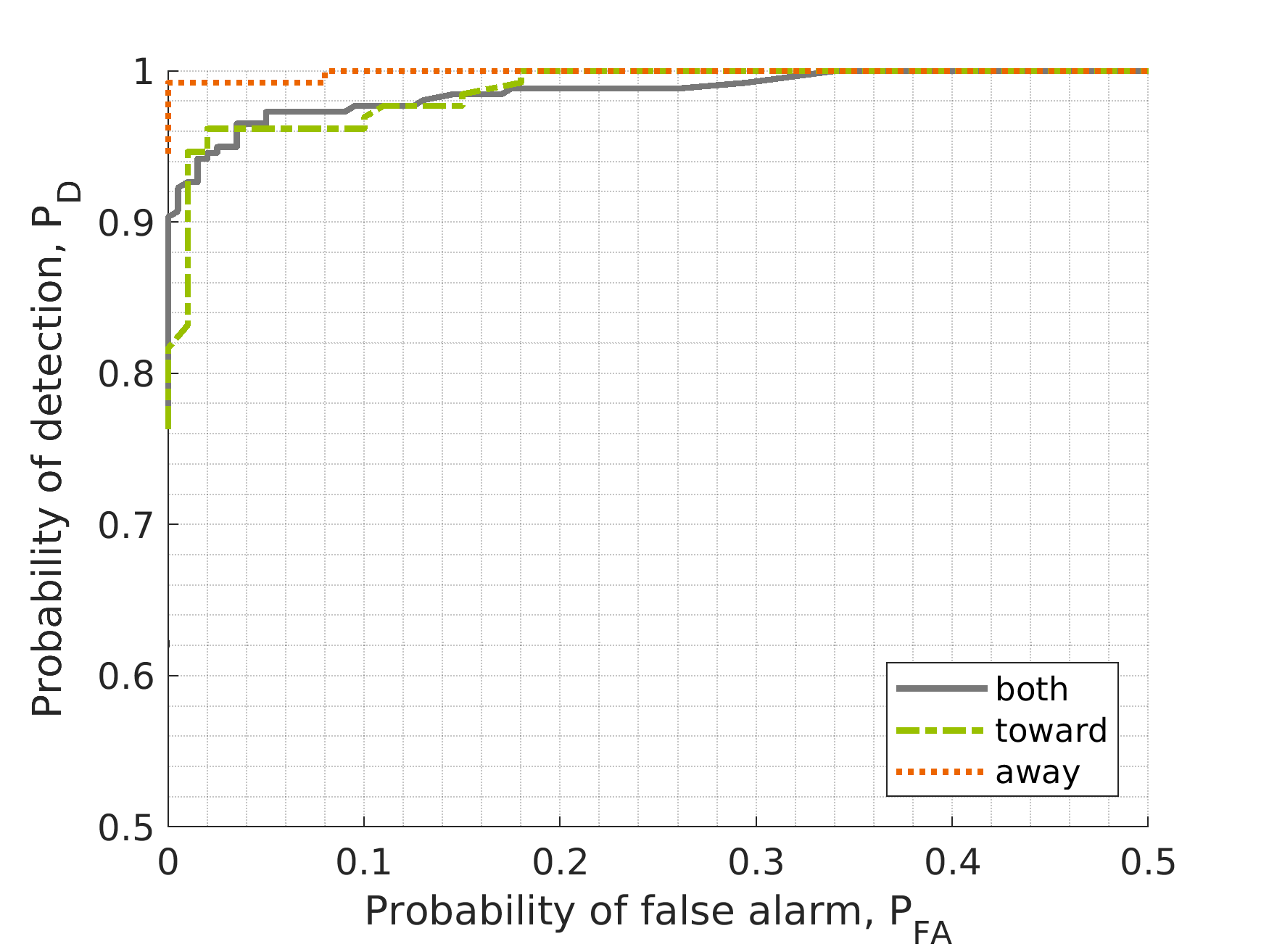}
	\caption{Receiver operating characteristic (ROC) for the best model per motion direction. In this case, Person M was excluded from the training set (see Table~\ref{tab:best_models}(c)).} 
	\label{fig:roc}
\end{figure}

Using the models given in Table~\ref{tab:best_models}, we calculate the probabilities of asymmetric gait $p(X)$ for Persons K--N. The threshold $\tau$ for deciding for or against asymmetric gait is chosen based on the training data such that the false alarm rate does not exceed 5\%, i.e., $P_{\text{FA}} \leq 5\%$. As an example, Fig.~\ref{fig:roc} shows the Receiver Operating Characteristic (ROC) of the regression model, which is trained on all available data except for data of Person M (see Table~\ref{tab:best_models}(c)). Observing the gait from the back ('away') results in very high detection rates even for small false alarm rates, whereas the scenarios 'toward' and 'both' are inferior.

Table~\ref{tab:pred_results} shows the probability of correctly detecting the asymmetric gait, $P_\text{D,test}$, of the four individuals with gait disorders. Additionally, the decision threshold $\tau$ and probability of detection on the training set, $P_\text{D,train}$, are given. In general, we can observe that the decision thresholds are lower for the 'away' scenario, which indicates that symmetric gait can more reliably be detected from behind (low false alarm rate). Thus, small values of $\tau$ can be chosen which increases $P_\text{D,train}$. Concerning $P_\text{D,test}$, we achieve very high rates in most of the cases, i.e., we can detect asymmetric gait with high probabilities. Although $P_\text{D,train}$ assumes higher values for the 'away' scenario throughout, we note that for some individuals, it is beneficial to monitor the gait from the front (see Persons L and M), whereas for Person K the 'away' scenario yields higher detection rates. Identifying the asymmetric gait of Person M in away-from-radar motions remains challenging, and $P_\text{D,test}$ assumes only 40\,\%.

\subsection{Discussion}
For in-home gait monitoring systems, the observation time is often limited owing to the inherent problem of short motion translation periods associated with household activities. Despite short observation times and using only four steps of the observed gait motion, we are able to detect the gait asymmetry of four diagnosed persons with high accuracy for at least one of the considered motion directions. Since the features are designed to detect differences \textit{between} the two leg motions, this is achieved irrespective of the degree of gait abnormality. Hence, the presented features are considered invariant to the actual appearance of the micro-Doppler step signatures, and thus, to different disorders. Clearly, data of more individuals is needed for a generalization of the obtained results.

\begin{table}[!t]
	\caption{Probability of detecting the asymmetric gait ($P_\text{D,test}$) of Persons K--N. The decision threshold $\tau$ is chosen based on the training data such that $P_\text{FA}\le 5\%$.}
	\centering
	\begin{tabular}{@{\extracolsep{2pt}} l r r r r r r}
		\toprule
		 & \multicolumn{3}{c}{Toward} & \multicolumn{3}{c}{Away} \\ 
		\cmidrule{2-4}\cmidrule{5-7} 
		 & $\tau$ & $P_\text{D,train}$ & $P_\text{D,test}$ & $\tau$ & $P_\text{D,train}$ & $P_\text{D,test}$ \\ 
		\cmidrule{1-7}
		K & $0.36$ & $96.95\,\%$ & $85.71\,\%$ & $0.18$ & $99.21\,\%$ & $\bm{100.00}\,\%$ \\
		L & $0.36$ & $96.06\,\%$ & $\bm{100.00}\,\%$ & $0.38$ & $97.58\,\%$ & $88.89\,\%$ \\
		M & $0.34$ & $96.18\,\%$ & $\bm{100.00}\,\%$ & $0.15$ & $99.22\,\%$ & $40.00\,\%$  \\	
		N & $0.35$ & $96.00\,\%$ & $\bm{100.00}\,\%$ & $0.17$ & $99.17\,\%$ & $\bm{100.00}\,\%$ \\
		\bottomrule
	\end{tabular}
	\label{tab:pred_results}
	\vspace{-0.8em}
\end{table}

\section{Conclusion}
\label{sec:conclusion}
Low-cost Doppler radar systems provide safe and privacy-preserving in-home sensing of human gait. It was shown that different gait disorders lead to distinct micro-Doppler signatures, which demonstrates the sensitivity of radar backscatterings to changes in gait. Asymmetric gait is quantified using image-based and physical features, which are extracted from the radar micro-Doppler step signatures represented in the joint Doppler frequency vs.~time domain. Based on real data measurements, including data of four persons with pathological gait, we were able to detect gait asymmetry with a high sensitivity, irrespective of the underlying gait disorder. Our results showed that, for radar-based gait analysis, it is beneficial to take the motion direction relative to the radar system into account.
  
\bibliographystyle{IEEEtran}
\bibliography{ref_radarconf19.bib}

\end{document}